\documentclass[useAMS,usegraphicx]{mn2e}
\usepackage{amsmath}
\usepackage{graphicx}

\title{The chemical evolution of galaxies with a variable IGIMF}

\author[Recchi \& Kroupa]{S.
  Recchi$^{1}$\thanks{simone.recchi@univie.ac.at} and P.
  Kroupa$^{2}$\thanks{pavel@astro.uni-bonn.de}\\ $^{1}$ Department of
  Astrophysics, Vienna University,
  T\"urkenschanzstrasse 17, A-1180, Vienna, Austria \\
  $^{2}$ Helmholtz-Institut f\"ur Strahlen- und Kernphysik (HISKP),
  Universit\"at Bonn, Rheinische Friedrich-Wilhelms-Universit\"at\\
  Nussallee 14-16 D-53115 Bonn Germany}
 
\date{Received; accepted}

\begin{document}
\maketitle


\begin{abstract}
  Standard analytical chemical evolution modelling of galaxies has
  been assuming the stellar initial mass function (IMF) to be
  invariant and fully sampled allowing fractions of massive stars to
  contribute even in dwarf galaxies with very low star formation rates
  (SFRs). Recent observations show the integrated galactic initial
  mass function (IGIMF) of stars, i.e. the galaxy-wide IMF, to become
  systematically top-heavy with increasing SFR. This has been
  predicted by the IGIMF theory, which is here used to develop the
  analytical theory of the chemical evolution of galaxies. This theory
  is non-linear and requires the iterative solution of implicit
  integral equations due to the dependence of the IGIMF on the
  metallicity and on the SFR. It is shown that the mass--metallicity
  relation of galaxies emerges naturally, although at low masses the
  theoretical predictions overestimate the observations by 0.3--0.4
  dex.  A good agreement with the observation can be obtained only if
  gas flows are taken into account.  In particular, we are able to
  reproduce the mass--metallicity relation observed by Lee et al.
  (2006) with modest amounts of infall and with an outflow rate which
  decreases as a function of the galactic mass.  The outflow rates
  required to fit the data are considerably smaller than required in
  models with invariant IMFs.
\end{abstract}


\maketitle


\section{Introduction}

The stellar initial mass function (IMF) has been traditionally
interpreted to be an invariant probability density distribution
function (e.g. Kroupa 2001, 2002; Bastian et al. 2010).  Recent
observational evidence has however ruled out this simple
interpretation in that low-density regions have been found to be
lacking the massive stars they ought to have produced given the large
number of stars formed (Kirk \& Myers 2011; Hsu et al.  2012). The
mass function of young star clusters has also been shown to not be a
probability density distribution function (Pflamm-Altenburg et al.
2013; Kroupa 2014).  Furthermore, several independent observational
evidences have been put forward in the last few years, showing that the
galaxy-wide IMF varies with star formation rate (SFR) (e.g.  Hoversten
\& Glazebrook 2008; Gunawardhana et al.  2011; Cappellari et al.
2012; Conroy \& van Dokkum 2012) and probably varies also within a
single galaxy (Dutton et al. 2013).

A comprehensive theory, attempting to explain the evolution of the IMF
within galaxies is the so-called integrated galactic IMF (IGIMF, i.e.
the galaxy-wide IMF) theory (Kroupa \& Weidner 2003; Weidner \& Kroupa
2005; see the recent review by Kroupa et al. 2013).  In two previous
papers (Recchi et al.  2009; Recchi et al. 2014) we have investigated
the effect of the IGIMF on the abundance ratios (global and of
individual stars) in galaxies.  One paper (Calura et al. 2010) has
been devoted to the study of the chemical evolution of the Solar
Neighborhood by adopting the (SFR-dependent) IGIMF formulation.
Eventually, numerical hydrodynamical simulations of dwarf galaxies
with a variable IGIMF have been pioneered, too (Ploeckinger et al.
2014).  It remains to be shown that the IGIMF theory is consistent
with the observed mass-metallicity (MZ) relation in galaxies (see e.g.
Lee et al. 2006; Kirby et al. 2013) and with the more general
fundamental metallicity relation (FMR, see Mannucci et al. 2010;
Lara-L{\'o}pez et al. 2010).  

We build on a previous attempt (K\"oppen et al. 2007) and solve the
equations of the so-called ``Simple Model'' but within the framework
of the IGIMF theory, i.e. we assume that the galaxy-wide IMF depends
on the SFR and on the metallicity of the parent galaxy. Analytical and
semi-analytical calculations of the chemical evolution of galaxies
based on simple models are nowadays quite popular, as they enlighten
in a simple way complex correlations among galaxies (Spitoni et al.
2010; Dayal et al. 2013; Lilly et al. 2013; Birrer et al. 2014; Pipino
et al. 2014).  It is worth remarking that all these authors consider
invariant IMFs whereas, as explained above, there is significant
observational evidence for IMF variations among different galaxies.
Therefore, it is important and timely to understand how to modify the
formalism of the simple model to take into account IMF variations.
This is indeed the main aim of the present paper.  It is also worth
stressing that, although this paper is based on the IGIMF theory,
other theories of IMF variation have been put forward in the
literature (e.g. Larson 1998; Padoan \& Nordlund 2002; Hopkins 2013.
See also Martinelli \& Matteucci 2000).  The formalism developed in
this paper can be applied equally well to these IMF theories.

The paper is organized as follows: in Sect. \ref{sec:igimf} we recall
the main assumptions of the IGIMF theory and describe the way in which
this theory has been implemented in the present work.  In Sect.
\ref{sec:simple} we describe the simple models of chemical evolution
-- a simplified tool to describe the evolution of the metallicity of a
galaxy.  In Sect. \ref{sec:cbrev} we re-interpret the so-called
closed-box model within the framework of the IGIMF theory, i.e. we
relax the hypothesis that the IMF is universal.  Similarly, in Sect.
\ref{sec:lbrev} we re-interpret simple models of chemical evolution
with outflows and infall and consider again variations of the IMF
between galaxies and within galaxies.  Finally, in Sec. \ref{sec:disc}
the results of this work are discussed and some conclusions are drawn.
In a companion paper (Recchi et al. 2014; hereafter Paper II), we will
apply this methodology to the study of the mass-metallicity relation
in tidal dwarf galaxies (TDGs).

\section{The IGIMF theory}
\label{sec:igimf}
The IGIMF theory is based on the assumption that stars in a galaxy are
formed in space-time correlated entities called embedded star clusters
(Lada \& Lada 2003), thereby accounting naturally for the above
mentioned results by Kirk \& Myers (2011), Hsu et al.  (2012) and
Pflamm-Altenburg et al. (2013). The notion underlying the IGIMF theory
lies in the conjecture that the freshly born galaxy-wide stellar
population is the sum of all local star-formation activity and that
this activity occurs in the density maxima of molecular cloud cores by
being made up of embedded groups or clusters of stars on a spacial
scale of less than a~pc (Marks \& Kroupa 2012).  The mass distribution
of these is equivalent to the mass function of embedded clusters which
do not need to be gravitationally bound upon expulsion of their
residual gas (Lada et al. 1984; Kroupa et al. 2001; Banerjee \& Kroupa
2013). An ``isolated'' or low-density mode of star formation is
naturally accounted for in this picture by low-mass and/or the
radially extended part of the embedded clusters.

In defining the IGIMF theory one posits the following six axioms
derived from observations (see Kroupa et al. 2013; Weidner et al.
2013)\footnote{Throughout this text $m$ and $M_{\rm ecl}$ are in units
  of $M_\odot$.}
\begin{enumerate}

\item The stellar IMF, $\xi(m)$, in an embedded star cluster is
  canonical (a two-part power law $\xi(m)\propto m^{-\alpha_i}$ with
  $\alpha_1=1.3$ for 0.08 M$_\odot$ $<m<$ 0.5 M$_\odot$, and
  $\alpha_2=2.3$ for stellar masses larger than 0.5 M$_\odot$) for
  cloud core densities $\rho_{\rm ecl} \leq 9.5 \times 10^4$ M$_\odot$
  pc$^{-3}$.  A star formation efficiency $\epsilon = M_{\rm
    ecl}/(M_{\rm ecl}+M_{\rm gas})$ in molecular cloud cores of 0.33
  has been assumed (Alves et al. 2007). Here $M_{\rm ecl}$ is the
  stellar mass in the embedded cluster and $M_{\rm gas}$ is the
  residual gas mass within it. 

\item The embedded clusters populate an embedded cluster mass function
  (ECMF), which is assumed to be a power law of the form $M_{\rm ecl}
  \propto M_{\rm ecl}^{-\beta}$ ($\beta=2.35$ would be the Salpeter
  index).

\item The half-mass radii of embedded clusters follow $r_{\rm h}$ (pc)
  = 0.1 $M_{\rm ecl}^{0.13}$ (Marks \& Kroupa 2012) yielding for the
  total density (gas plus stars) log$_{10}$($\rho_{\rm cl}$) = 0.61
  log$_{10}$($M_{\rm ecl}$) + 2.85, in units of M$_\odot$ pc$^{-3}$.

\item The most massive star in a cluster, $m_{\rm max}$, is a function
  of the stellar mass of the embedded cluster, $M_{\rm ecl}$ (Weidner
  \& Kroupa 2004, 2006; Weidner et al. 2010), $m_{\rm max}=m_{\rm max}
  (M_{\rm ecl})$ $\le m_{\rm max*}\approx 150\,M_\odot$ (e.g. eq. 10
  in Pflamm-Altenburg et al. 2007).

\item There exists a relation between the SFR of a galaxy and the most
  massive young (age$<$10 Myr) star cluster, log$_{10}$($M_{\rm ecl}$)
  = 0.764 log$_{10}$(SFR)+4.93 (Weidner et al. 2004), where the SFR is
  in units of M$_\odot$ yr$^{-1}$.  The mass of the least-massive
  embedded ``cluster'' is assumed to be 5 M$_\odot$.

\item The dependence of the IMF slope, $\alpha_3$, of stars above 1
  M$_\odot$ on the initial density and metallicity of the embedded
  cluster is given by (Marks et al. 2012)
\begin{equation}
\alpha_3=\begin{cases}2.3&\mbox{if}\; x<-0.87\\
-0.41 \times x + 1.94 & \mbox{if}\; x\geq -0.87\end{cases},
\label{eq:alpha3}
\end{equation}
\noindent
where 
\begin{equation}
x=-0.14{\rm [Fe/H]}\,+0.99 \log_{10}\left(\frac{\rho_{\rm cl}}{10^6 
{\rm M}_{\odot} {\rm pc}^{-3}}\right).
\end{equation}
\noindent
This dependency of the stellar IMF on density and metallicity of a
massive star-burst cluster has been derived by combining independent
evidence from globular clusters (Marks et al. 2012) and ultra-compact
dwarf galaxies (Dabringhausen et al. 2009, 2012). 
\end{enumerate}

For a value of $\beta=2$ the IGIMF built in this way nicely reproduces
many observed properties of galaxies (see Kroupa et al. 2013).  It
also qualitatively reproduces the trend of decreasing $\alpha_3$ with
increasing SFR observed by Gunawardhana et al. (2011).   In order to
improve this fit, a seventh axiom has been introduced (Weidner et al.
2013): the exponent $\beta$ of the ECMF is not a constant but it
varies with the SFR according to:
\begin{equation}
\beta=\begin{cases}2.00&\mbox{if}\; SFR<1\;{\rm M}_\odot 
{\rm yr}^{-1},\\
-0.106\log_{10}({\rm SFR})+2.00 &\mbox{if}\; SFR \geq 1\;{\rm M}_\odot 
{\rm yr}^{-1}.\end{cases}
\label{eq:betad}
\end{equation}
\noindent
Axiom~7 implies the mass function of embedded clusters to be top-heavy
in star bursts.  Weidner et al. (2013) explored also the possibility
that the mass of the smallest embedded cluster depends on the SFR,
thus partially relaxing the fifth axiom above.  They showed however
that this assumption produces results almost indistinguishable from
the results obtained with axiom 7, thus we will not further consider
this hypothesis.

\section{Simple models of chemical evolution: 
the invariant IMF case}
\label{sec:simple}
A class of galactic chemical evolution models, commonly dubbed as
Simple Models, admit analytical solutions.  They are thus quite useful
and easy-to-apply diagnostics for the evolution of galaxies (see
Spitoni et al.  2010; Lilly et al. 2013; Pipino et al. 2014; see also
the Introduction), thus it is useful to understand advantages and
limitations of this formalism.

The basic assumptions behind the simple models are (see Tinsley 1980;
Maeder 1992):
\begin{enumerate}
\item Stars more massive than a fixed mass (usually 1 M$_\odot$) die
  instantaneously.  Stars less massive than this threshold live
  forever.  This assumption (instantaneous recycling approximation)
  allows us to neglect the lifetime of single stars in what follows.
\item The gas is well mixed at any time.
\item The flow rates (infall and outflow) are proportional to the SFR
  $\psi$.
\item The IMF is universal: it is independent of time and it is the
  same in all galaxies.
\item The yields and the remnant masses are functions of the stellar
  mass alone.  This assumption (together with the universality of the
  IMF) implies that the quantities:
  \begin{align}
    &R=\int_1^\infty \left(m-m_{\rm R}\right)\xi(m)dm,\notag\\
    &y_{\rm Z}=\frac{1}{1-R}\int_1^\infty m \, p_{{\rm Z} m} \, \xi(m)dm,
    \label{eq:ryz}
  \end{align}
  \noindent
  are constant.  Here, $\xi$ is the adopted (fixed, for the moment)
  IMF, $m_{\rm R}$ is the remnant mass of the star with initial mass
  $m$, and $p_{{\rm Z} m}$ is the mass-fraction of newly synthesised
  metals by the star of initial mass $m$.  According to this
  terminology, the total amount of heavy elements released by a star
  of initial mass $m$, initial metallicity $Z$ and remnant mass
  $m_{\rm R}$ is $m_{\rm ej, Z}=mp_{{\rm Z} m}+(m-m_{\rm R})Z$, i.e.
  it includes newly synthesised heavy elements but also metals present
  in the stars at birth and not processed.  The quantity $R$ in Eq.
  \ref{eq:ryz} represents thus the fraction of a stellar population
  not locked into long-living (dark) remnants and $y_{\rm Z}$
  represents the ratio between the mass of heavy elements ejected by a
  stellar generation and the mass locked up in remnants.
\end{enumerate}

It is well known that these assumptions lead to a set of differential
equations for the time evolution of the total baryonic mass $M_t$,
total gas mass $M_g$ and total mass in metals $M_Z$ within a galaxy
(see Tinsley 1980; Maeder 1992; Pagel 1997; Matteucci 2001):
\begin{equation}
  \begin{cases}{d M_{t} \over d t} = (\Lambda - \lambda) (1 - R) \psi (t) \cr
    {d M_g \over d t} = (\Lambda - \lambda - 1) (1 - R) \psi (t) \cr
    {d M_Z \over d t} = (1 - R) \psi (t) [\Lambda Z_A + y_Z - 
    (\lambda + 1) Z]\end{cases}
\label{eq:system}
\end{equation}
\noindent
Here, $Z_A$ is the metallicity of the infalling material, whereas
$\Lambda$ and $\lambda$ are proportionality constants relating the SFR
to the infall and outflow rate, respectively.  In particular, it is
assumed that:
\begin{equation}
A(t)=\Lambda (1-R)\psi(t),\;\;\;\;W(t)=\lambda(1-R)\psi(t),
\label{eq:flows}
\end{equation}
\noindent (see Matteucci \& Chiosi 1983) where $A(t)$ and $W(t)$ are
the infall or accretion and the outflow rate, respectively.  Given the
initial condition $M_t(0)=M_g(0)=M_0$, $Z(0)=Z_0$, a solution as a
function of the mass ratio $\mu=M_g/M_t$ can be found (see also Recchi
et al. 2008):
\begin{equation}
  Z=Z_0+\frac{\Lambda Z_A +y_{\rm Z}}{\Lambda}\left\lbrace
    1-\left[(\Lambda-\lambda)-\frac{\Lambda-\lambda-1}{\mu}\right]^{
      \frac{\Lambda}{\Lambda-\lambda-1}}\right\rbrace.
\label{eq:simplechevsol}
\end{equation}
\noindent
This solution expresses the gas-phase metallicity of a galaxy, as a
function of the mass fraction $\mu$.  The time dependence of this
relation is hidden in the time dependence of $\mu$, which in turn
depends on how quickly the gas is consumed to form stars (i.e. it
depends on the SFR) and on the flows in and out of the galaxy.  Hence,
to have a complete picture of the evolution of the galaxy, we need
also some assumption concerning the SFR.  In what follows we will
assume that the SFR depends on the available gas reservoir through the
simple law
\begin{equation}
\psi = s \cdot M_g,\;\;\;\;\;s=0.3\,{\rm Gyr}^{-1},
\label{eq:sflaw}
\end{equation}
\noindent
i.e. we assume here that the SFR is linearly proportional to the gas
mass and that this proportionality constant is an invariant quantity.
This equation also implies that the gas consumption timescale of a
galaxy is always about 3 Gyr.  Notice that many studies of the
chemical evolution of galaxies suggest that the star formation
efficiency increases with the mass of the galaxy (e.g. Matteucci 1994;
Pipino \& Matteucci 2004).  These studies are however based on
invariant IMFs and indeed Matteucci (1994) showed that a variable IMF
or a variable star formation efficiency might produce similar results.
In the framework of the IGIMF, Pflamm-Altenburg \& Kroupa (2009)
showed that the gas consumption time-scale is similar in all galaxies,
irrespective of their masses, and it is indeed of the order of 3 Gyr.
We consider also the average metallicity $Z_*$ of the stars in a
galaxy.  A simple way of calculating this quantity is:
\begin{equation}
Z_*=\frac{\int_0^t Z (t) \psi (t)dt}{\int_0^t \psi(t)dt}.
\label{eq:zstar}
\end{equation}
\noindent
This expression represents the mass-weighted average of the
metallicities of all the stellar populations ever born in the galaxy
(see Pagel 1997).  A more accurate definition of $Z_*$ would involve a
luminosity-weighted average, but it has been shown that these two
averages do not differ much (e.g. Recchi et al. 2009).

A solution can be found also in the presence of metal-enhanced
galactic outflows (see Sect. \ref{sec:lbrev}).  A solution can also be
obtained assuming generic infall and outflow laws (see again Recchi et
al. 2008).  From the given solution Eq. \ref{eq:simplechevsol}, other
particular cases can be found.  For instance, setting
$\Lambda=\lambda=\alpha=Z_A=0$ one obtains the well-known solution of
the closed-box model $Z=-y_{\rm Z} \ln (\mu)$\footnote{In Eq.
    \ref{eq:simplechevsol} a $\Lambda$ is in the denominator, hence
    $\frac{\Lambda Z_A +y_{\rm Z}}{\Lambda}$ diverges as $\Lambda
    \rightarrow 0$.  The previous sentence might thus look suspicious.
    The fact is that also the expression within the curly brackets
    tends to zero for $\Lambda \rightarrow 0$.  Take $\lambda=Z_A=0$;
    the solution reduces to
    \begin{equation}Z=\frac{y_{\rm Z}}{\Lambda}\left[ 1 -
        \left(\Lambda
          -\frac{\Lambda-1}{\mu}\right)^{\frac{\Lambda}{\Lambda-1}}\right]
      \simeq \frac{y_{\rm Z}}{\Lambda}\left[ 1 -
        \left(\frac{1}{\mu}\right)^{-\Lambda}\right],\label{eq:foot}\end{equation}
    \noindent
    where we have taken the limit for small $\Lambda$ in the second
    equality.  Write $\left(\frac{1}{\mu}\right)^{-\Lambda}$ as
    $e^{-\Lambda \ln (1/\mu)}$ and expand it in a Taylor series, e.g.
    $e^{-\Lambda \ln (1/\mu)}\simeq 1- \Lambda \ln (1/\mu)$.  By
    substituting this expansion in the expression for $Z$ in Eq.
    \ref{eq:foot}, the closed box solution $Z=y_{\rm Z}\ln(1/\mu)$ is
    recovered.}

It is generally accepted that this solution gives a first-order
accurate description of the overall evolution of chemical elements
(such as the $\alpha$-elements) mainly released by massive stars on
short timescales.

\section{The closed-box model revisited: the variable IGIMF}
\label{sec:cbrev}
The relevant equations for the closed-box model with a linear SFR such
as in Eq. \ref{eq:sflaw} are:
\begin{equation}
  \begin{cases}{d M_{t} \over d t} = 0 \cr
    {d M_g \over d t} = - (1 - R) \psi (t) \cr
    {d M_Z \over d t} = (1 - R) \psi (t) [y_Z - Z] \cr
    \psi (t) = s M_g (t) \end{cases}
\label{eq:cbsystem}
\end{equation}
\noindent
It is trivial to solve for $Z(t)$, $M_g(t)$.  Assuming that there is
only gas at the beginning and that $Z(0)=0$, one obtains:
\begin{equation}
Z(t)=s(1-R)y_Zt,\;\;\;M_g(t)=M_g(0)\exp{[-s(1-R)t]}.
\label{eq:solcb}
\end{equation}
\noindent
It is important to remark on this equation that, according to the
assumptions of the simple model, all constants that enter in the
calculation of $Z(t)$ are indeed constants, i.e. they do not depend on
the mass of the model galaxy taken into consideration.  The
consequence is that the time evolution of the metallicity is always
the same, independent of the mass of the galaxy.  Consequently, these
models can not reproduce the observed MZ relation.  A viable solution
is to assume that small galaxies can experience galactic winds, due to
their shallow potential wells.  Observations show indeed that
mass-loading factors in galactic winds are larger in small galaxies
(Martin 2005) and theoretical investigations confirm this finding
(Hopkins et al. 2012; Puchwein \& Springel 2013). If the parameter
$\lambda$ introduced in Sect. \ref{sec:simple} is large in small
galaxies and small or zero in large ones, the MZ relation can be
nicely reproduced (Tremonti et al.  2004; Spitoni et al. 2010). 
 
Although we are convinced that galactic winds can play an important
role in the chemical evolution of dwarf galaxies (Recchi \& Hensler
2013; see also Sect. \ref{sec:lbrev}), we want to explore here the
possibility that the IMF is not constant but that it depends on time
though its dependence on the SFR and on Z, as per the IGIMF theory
described in Sect. \ref{sec:igimf}.  As explained in the Introduction,
this exploration is purely didactic and serves to illustrate in the
simplest possible setting how a variable IMF affects the formalism of
the simple model described in Sect.  \ref{sec:simple}.

Clearly, the assumption of a variable IMF introduces a differentiation
in the metallicity evolution of model galaxies with different masses,
as the SFR of different model galaxies are different and that implies
variations in the IGIMF, which in turn imply variations in the
chemical evolution.

To be more quantitative, the quantities $R$ and $y_Z$ introduced in
Eq. \ref{eq:ryz} become:

\begin{align}
  &R(t)=\int_1^\infty \left(m-m_{\rm R}\right)\xi_{\rm IGIMF}
  \left[m,\psi(t),Z(t)\right]dm,\notag\\
  &y_{\rm Z}(t)=\frac{1}{1-R}\int_1^\infty m \, p_{{\rm Z} m} \, \xi_{\rm IGIMF}
  \left[m,\psi(t),Z(t)\right]dm,
  \label{eq:ryz2}
\end{align}
\noindent
i.e. they are not constant any more, but are functions of time,
through the time dependence of $\psi$ and $Z$.  The IMF $\xi_{\rm
  IGIMF}$ entering here is the IGIMF, i.e. the IMF obtained adopting
the axioms described in Sect. \ref{sec:igimf}.  The formal solution of
the closed-box equations \ref{eq:cbsystem} is now:
\begin{align}
  &Z(t)=Z_0+s\int_0^t \left[1-R(\tau)\right]y_Z(\tau)d\tau, \notag\\
  &M_g(t)=M_g(0)\exp{\left[-s\int_0^t \left[1-R(\tau)\right]d\tau\right]},
  \notag\\
  &\psi(t)=sM_g(t)=sM_g(0)\exp{\left[-s\int_0^t
    \left[1-R(\tau)\right]d\tau\right]}.
\label{eq:sol1}
\end{align}
\noindent
We have included also the solution for $\psi(t)$ to show that this
system of equations is indeed implicit (the integrals in the right
hand sides depend on the quantities on the left hand sides).  To solve
these equations thus, an implicit procedure must be employed.

\begin{figure*}
\resizebox{8cm}{!}{\includegraphics[angle=270]{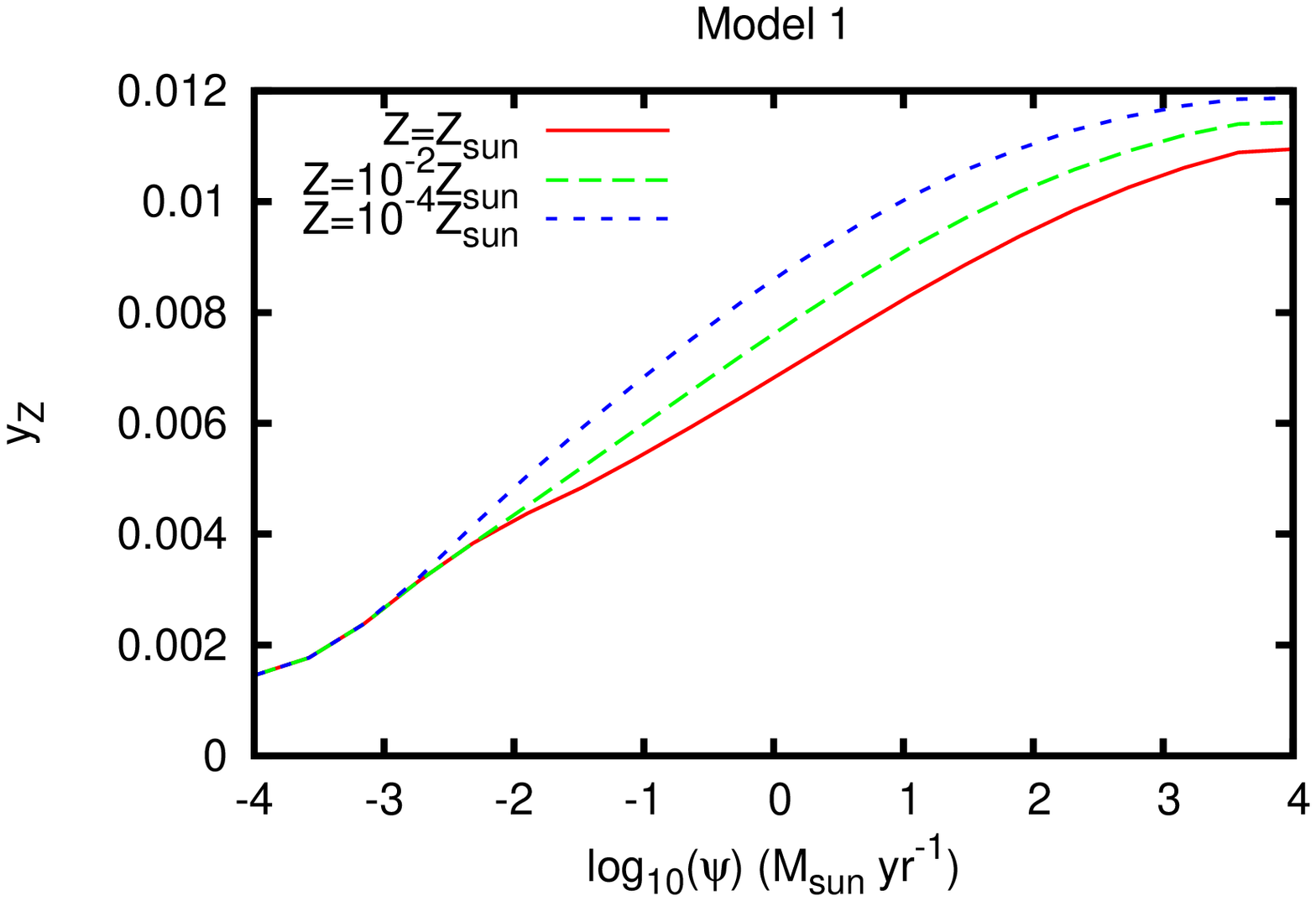}}
\resizebox{8cm}{!}{\includegraphics[angle=270]{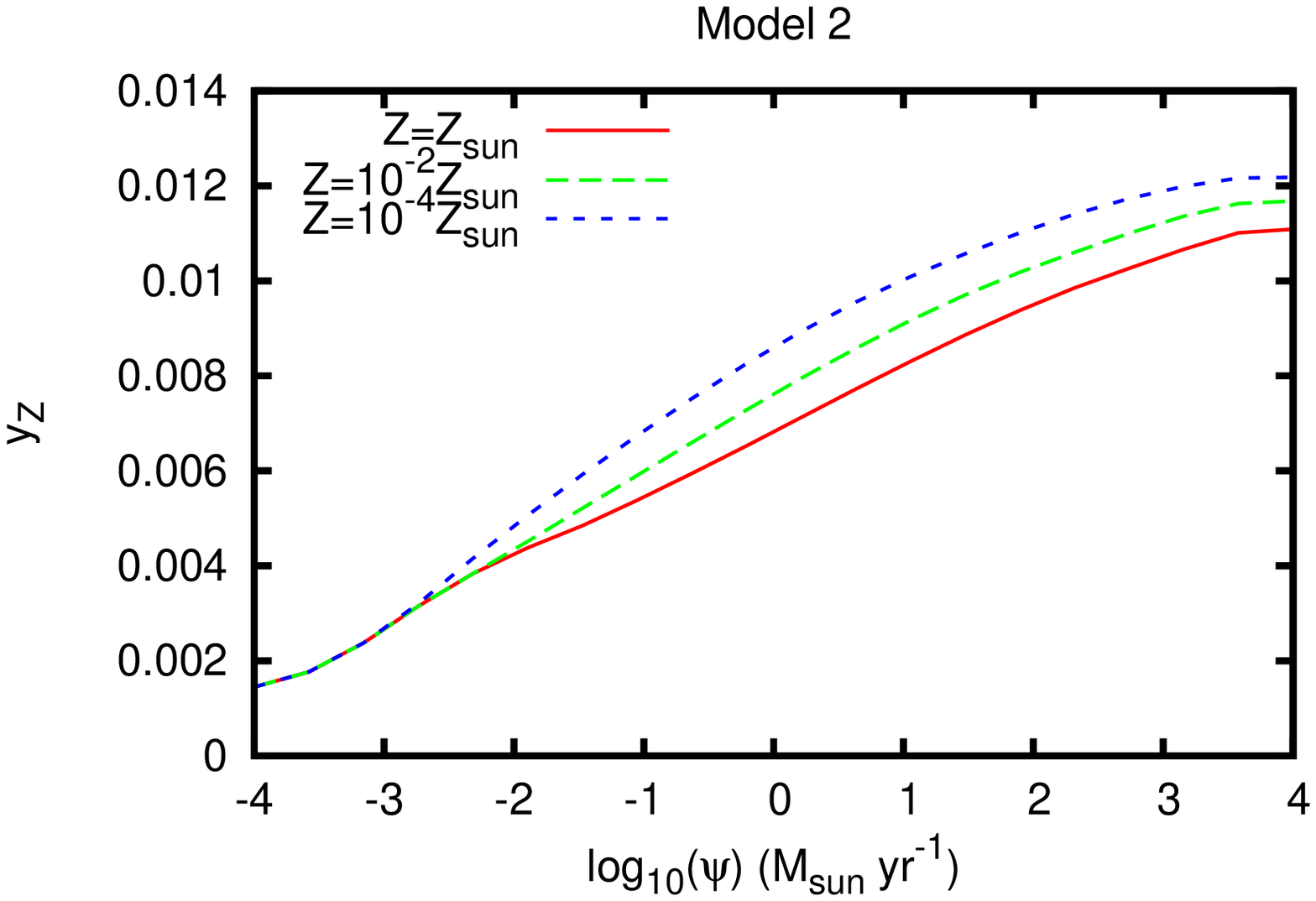}}
\caption{The yield $y_Z$, as defined in Eq. \ref{eq:ryz2}, as a
  function of the SFR $\psi$ and of the metallicity.  The left panel
  refers to the Model 1 (axioms 1--6 of Sect. \ref{sec:igimf}) and the
  right panel refers to Model 2 (where also the last axiom of Sect.
  \ref{sec:igimf} is implemented). In comparison, an invariant IMF has
  $y_Z=$constant (Sec.~\ref{sec:simple}).}
\label{fig:y_z}
\end{figure*}
Examples of the calculation of the yields $y_Z$ using the IGIMF theory
and the axioms described in Sect.  \ref{sec:igimf} are shown in
Fig.~\ref{fig:y_z}.  In particular, the left panel refers to a model
(Model 1) in which the axioms 1--6 of Sect. \ref{sec:igimf} are
adopted.  The right panel instead refers to Model 2, i.e. a model
where also the last axiom of Sect.  \ref{sec:igimf} is implemented.
As mentioned in the Introduction, the methodology described in this
paper is not limited to the IGIMF theory.  Other theories of variable
IMF could be used to calculate $y_Z$ and $R$ as a function of galactic
properties.  Once $y_Z$ and $R$ have been calculated, the rest of the
theory is general and not related to a specific IMF formulation.

To construct these functions $y_Z(t)$, we have adopted the yields of
Woosley \& Weaver (1995) for massive stars and the ones of van den
Hoek \& Groenewegen (1997) for intermediate-mass stars.  Both sets of
yields are metallicity-dependent.  Since we are interested in the
early phases of the evolution of galaxies, we have thus taken the set
of yields with Z=0.001 from van den Hoek \& Groenewegen (1997) and the
one with Z=10$^{-4}$ Z$_\odot$ from Woosley \& Weaver (1995).  We have
checked that our results depend little on the chosen metallicity,
although larger initial metallicities tend to produce slightly higher
yields.  The increase of $y_Z$ with SFR is due to the fact that the
IGIMF is flatter for high SFRs, thus more metals are produced.
Because of Eq.  \ref{eq:alpha3} (and because of the dependence of $x$
on [Fe/H]), the yield increases also with decreasing metallicity. The
yields of Model 2 are slightly larger than the yields of Model 1,
because of the axiom 7 of Sect. \ref{sec:igimf}.  The differences
between Model 1 and Model 2 however disappear for SFRs smaller than 1
M$_\odot$ yr$^{-1}$ because of Eq. \ref{eq:betad}.  Given the small
differences between Model 1 and Model 2, we will focus from now on on
results pertaining Model 1.  Notice also that the yields tend to
flatten out at very low SFRs.  This is due to the fact that low- and
intermediate-mass AGB stars still produce some metals according to the
yields of van den Hoek \& Groenewegen and the fraction of AGB stars
(in particular the fraction of low-mass AGB stars) changes very little
with the SFR.  The function $R(\psi)$ is qualitatively similar to the
shown function $y_Z(\psi)$, although the magnitude of the variations
is smaller.

\begin{figure*}
\resizebox{8cm}{!}{\includegraphics[angle=270]{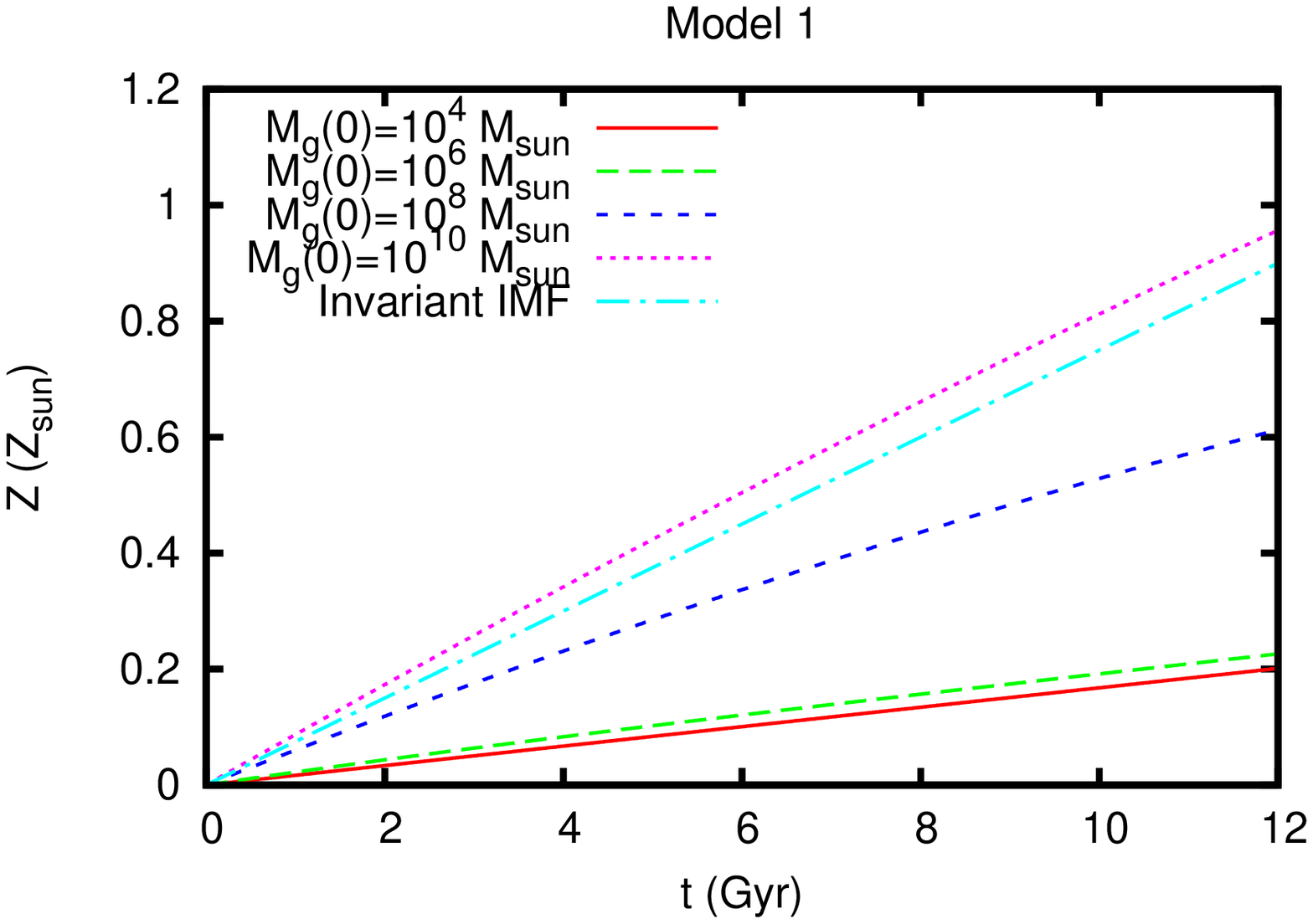}}
\resizebox{8cm}{!}{\includegraphics[angle=270]{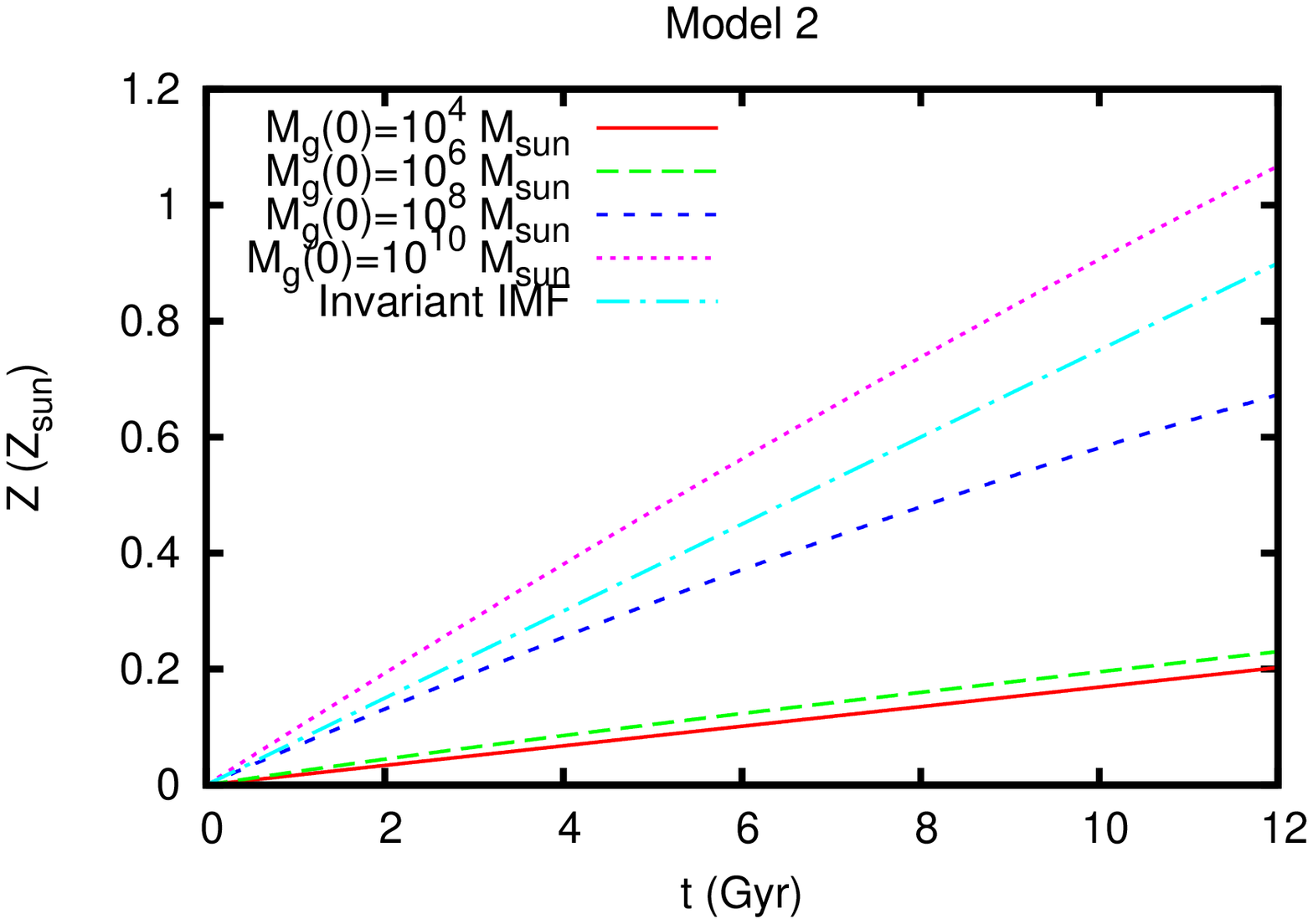}}
\caption{The gas-phase metallicity $Z$ as a function of time for four
  representative model galaxies with different initial gas masses,
  from 10$^4$ M$_\odot$ (lower curves) to 10$^{10}$ M$_\odot$ (upper
  curves).  The left panel refers to Model 1 (axioms 1--6 of Sect.
  \ref{sec:igimf}) and the right panel refers to Model 2 (where also
  the last axiom of Sect.  \ref{sec:igimf} is implemented).  In each
  panel a model with an invariant IMF (thus with no dependence on the
  initial gas mass of the model) is shown for comparison.}
\label{fig:traces}
\end{figure*}

Fig. \ref{fig:traces} depicts the time evolution of four
representative model galaxies with different initial gaseous masses,
ranging from 10$^4$ M$_\odot$ (lower curves) to 10$^{10}$ M$_\odot$
(upper curves).  The left panel refers to the Model 1 (axioms 1--6 of
Sect.  \ref{sec:igimf}) and the right panel refers to Model 2 (where
also the last axiom of Sect.  \ref{sec:igimf} is implemented).  In
these models it is assumed that all the mass initially is in the form
of gas.  Clearly, the metallicity evolution depends now, at variance 
with the invariant IMF case, on the model galaxy, as larger galaxies
have on average higher SFR, hence flatter IMFs, with a higher
production rate of heavy elements.  As a comparison, a model with an
invariant IMF is shown in Fig.  \ref{fig:traces}, too.  Typical
parameters ($y_Z=0.01$ and $R=0.5$) have been used to plot these
curves.  It is true that low-metallicity models tend to have flatter
IMFs, too (see Eq.  \ref{eq:alpha3} and the definition of the
parameter $x$), but since the proportionality constant in front of
[Fe/H] in the definition of $x$ is small, the metallicity effect on
the IMF is outweighted by the SFR effect.  Due to the larger yields,
the final metallicities for Model 2 are (slightly) larger than the
ones for Model 1.

\begin{figure}
  \resizebox{8cm}{!}{\includegraphics[angle=270]{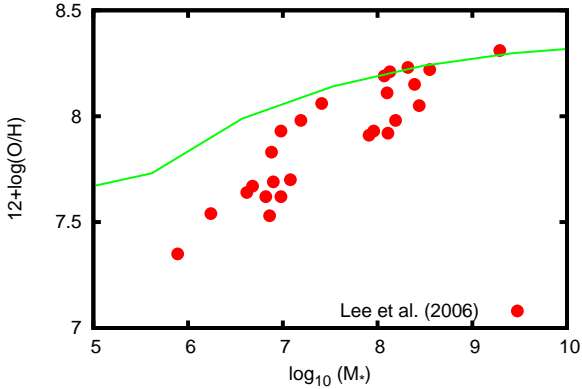}}
  \caption{The MZ relation obtained at an age of 12~Gyr by means of
    the simple closed box model within the IGIMF theory adopting Model
    1 (solid line; Model~2 is barely distinguishable from Model~1 as
    is evident in Fig.~\ref{fig:traces}.). The Lee et al. (2006)
    observations are shown as red circles.}
\label{fig:mz}
\end{figure}

For all these model galaxies, we have also calculated the average
stellar metallicities, according to Eq. \ref{eq:zstar}.  In principle,
these stellar metallicities could be compared with the best
observational data available to us, i.e. the stellar metallicities of
dwarf satellites of the Milky Way and in general of dwarf galaxies of
the Local Group.  The stellar metallicities of these galaxies are well
determined and they build a clear and tight MZ relation (see e.g.
Kirby et al. 2013).  Unfortunately, the [Fe/H] abundance is usually
measured in stars (see again Kirby et al.  2013).  This makes the
comparison with simple model predictions much less straightforward.
In fact, as it is well known, the majority of iron is produced by Type
Ia Supernovae, over long timescales. As recalled in Sect.
\ref{sec:simple}, the simple models give acceptable results only for
elements (such as the $\alpha$-elements) produced on short timescales
by massive stars.  Hence, simple models adopting the instantaneous
recycling approximation cannot predict the right Fe abundance since
they cannot take into account properly the contribution from Type Ia
SNe.  We are working on an extension of the simple model formalism
that takes into account also iron produced by SNeIa.  At the moment,
results of simple models can be safely compared to gas-phase
abundances of $\alpha$-elements only, and this is what we do in Fig.
\ref{fig:mz}.  We use the data of Lee et al.  (2006) and we compare
them with the results of model galaxies shown in Fig.
\ref{fig:traces}.  The initial metallicity considered here (the $Z_0$
in Eq. \ref{eq:sol1}) is zero. Initial metallicities larger than zero
are important when studying TDGs, therefore they will be considered
extensively in Paper II. We anticipate however here that, if TDGs form
at very large redshifts out of galaxies whose chemical enrichment is
still limited, their initial metallicity is not large and $Z_0=0$ is a
reasonable assumption.

Clearly, we can see a MZ in Fig. \ref{fig:mz}, in the sense that
low-mass models attain smaller stellar metallicities than high-mass
ones.  However, the comparison with observations is not particularly
good.  In particular at low masses the theoretical predictions
overestimate the observations by 0.3--0.4 dex.  This is due in part to
the inherent uncertainties in the Lee et al. (2006) MZ relation, in
part to the fact that galactic flows in low-mass systems can not be
neglected.  We analyze in detain the effect of galactic flows in the
following section.

\section{The leaky box model revisited}
\label{sec:lbrev}
We revise now the system of equations \ref{eq:cbsystem} to take into
account possible gas inflows and outflows in galaxies. As already
mentioned, this will be done by introducing parameters $\Lambda$ and
$\lambda$ that relate the flow rates to the SFR (see Eq.
\ref{eq:flows}).  Additionally, we consider the possibility that the
outflows are metal-enriched ({\it differential winds}), i.e. that the
metallicity of the outflowing gas is larger than the average gas-phase
metallicity in the galaxy.  Both observations (Martin et al. 2002; Ott
et al. 2005) and numerical models (Mac-Low \& Ferrara 1999; Recchi \&
Hensler 2013) support this assumption.  Notice that probably also the
efficiency with which the various chemical species are ejected vary
from element to element ({\it selective winds}; see Recchi et al.
2001; 2008) but for simplicity we will not consider further this
hypothesis.

Assuming that the metallicity of the outflowing gas is $Z_o=\gamma Z$,
with $\gamma \geq 1$, we can write:

\begin{equation}
  \begin{cases}{d M_{t} \over d t} = (\Lambda-\lambda)(1-R)\psi(t), \cr
    {d M_g \over d t} = (\Lambda- \lambda - 1)(1 - R) \psi (t), \cr
    {d M_Z \over d t} = (1 - R) \psi (t) [\Lambda Z_A + 
    y_Z - (\gamma \lambda +1)Z], \cr
    \psi (t) = s M_g (t). \end{cases}
\label{eq:lbsystem}
\end{equation}
\noindent
The solution of this system of equations for $M_g(t)$ and $Z(t)$ is
given by:
\begin{align}
  &M_g(t)=M_g(0)\exp{\left[(\Lambda-\lambda-1)s\int_0^t
      \left[1-R(\tau)\right]d\tau\right]},
  \notag\\
  &Z(t)=Z_0+\frac{\int_0^t[1-R(\tau)]s[\Lambda Z_A + y_Z(\tau)]
    I(\tau)d\tau}{I(t)}, \notag\\
  &I(t)=\exp{\left[-\int_0^t[1-R(\tau)]s[\lambda(\gamma-1)+\Lambda]
      d\tau. \right]}
  \label{eq:sol2}
\end{align}
\noindent
Here, $I(t)$ is an appropriate integration factor.  Notice that the
expression for $Z(t)$ reduces to the analogous closed-box expression
shown in Eq. \ref{eq:sol1} in the case $\gamma=1$, $\Lambda=0$ (i.e.
no differential winds, no infall).

\begin{figure}
\resizebox{8cm}{!}{\includegraphics[angle=270]{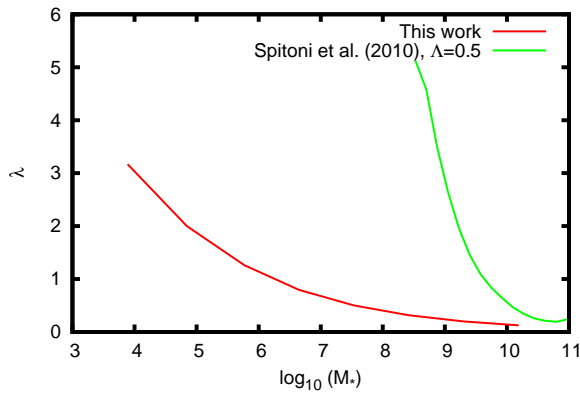}}
\caption{The comparison between the outflow rates $\lambda$ adopted in
  this study (see Eq. \ref{eq:mglambda}) and the ones adopted for
  $\Lambda=0.5$ in Spitoni et al. (2010).}
\label{fig:lambda_comp}
\end{figure}

As already mentioned, it is to be expected that low-mass galaxies are
much more affected by galactic winds than are more-massive ones, i.e.
we expect $\lambda$ to decrease as the galactic mass increases.  A
simple functional form satisfying this condition is\footnote{Actually,
  one expects $\lambda$ to decrease with increasing total (baryons
  plus dark matter) galactic mass.  If we assume for simplicity a
  constant ratio between dark matter and baryonic mass, a correlation
  between $\lambda$ and $M_g(0)$ as in Eq. \ref{eq:mglambda} is
  recovered.}:
\begin{equation}
\lambda=2 \cdot 10^{1-0.2 \log_{10}M_{g}(0)}.
\label{eq:mglambda}
\end{equation}
\noindent
This formula is normalized in order to reproduce the outflow rate to
SFR ratio in TDG simulations in which the IGIMF prescriptions are
adopted (Ploeckinger et al. 2014). According to this formula,
$\lambda$ tends to be quite large for very low mass systems (it is
$\lambda \simeq 3.17$ for $M_g(0)=10^4$ M$_\odot$) but it tends to
zero for very high mass ones (it is $\lambda \simeq 0.13$ for
$M_g(0)=10^{11}$ M$_\odot$).  On the other hand, we do not expect the
infall to significantly depend on the mass of the parent galaxy,
therefore we take, in compliance with Spitoni et al. (2010),
$\Lambda=0.5$ for all model galaxies.  Notice that the values of
$\lambda$ are considerably smaller than the ones needed in order to
reproduce the MZ relation based on an invariant IMF. The comparison
between our adopted values of $\lambda$ and the ones of Spitoni et al.
(2010) for $\Lambda=0.5$ are given in Fig.  \ref{fig:lambda_comp}.
Notice that we are interested in this study in smaller systems
compared to the ones considered by Spitoni et al.  (2010), therefore
the two curves in Fig. \ref{fig:lambda_comp} do not fully overlap.  It
is nevertheless clear that the outflow rates in Spitoni et al. (2010)
are much larger than the ones adopted in this study.  This is due to
the fact that the reduction of the metallicity at low masses is,
according to the IGIMF theory, largely due to the steepening of the
IMF slope at large stellar masses in galaxies with small SFRs.
Moreover, we assume $\gamma=2.5$, based on the results of the
chemo-dynamical simulations of Recchi \& Hensler (2013).

\begin{figure}
\resizebox{8cm}{!}{\includegraphics[angle=270]{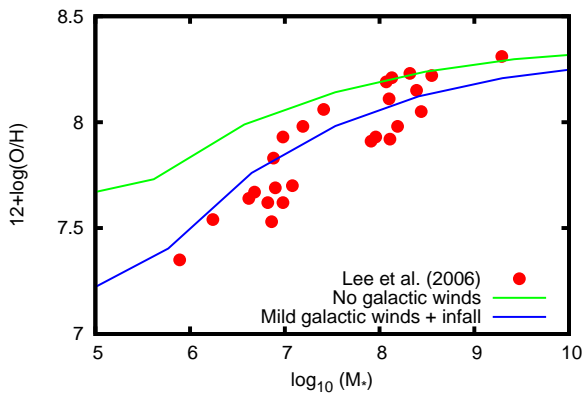}}
\caption{The MZ relation obtained by means of the leaky box model
  within the IGIMF theory.  $\Lambda$ is assumed to be equal to 0.5
  and the correlation between $\lambda$ and $M_g(0)$ is as in Eq.
  \ref{eq:mglambda}.}
\label{fig:mz_w}
\end{figure}

The MZ relation obtained with leaky box models with this dependence of
$\lambda$ with $M_g(0)$ is shown in Fig.  \ref{fig:mz_w}.  The fit
with observations is very good.  It could be better if we better
fine-tune the relation $\lambda=\lambda(M_g(0))$ but this is not the
main aim of the paper.

\section{Discussion and conclusions}
\label{sec:disc}

The IGIMF theory (Kroupa et al. 2013) predicts a coupling between some
properties of a galaxy (the SFR and the metallicity) and the IMF.
Since the IMF in turn strongly affects the dynamical evolution of the
galaxy, the feedback between galaxy evolution and IMF is difficult and
the fully complexity of a variable IMF has not been yet included in
hydrodynamical simulations (but see Bekki 2013; Ploeckinger et al.
2014; Recchi 2014). Even the here treated approach based on the
so-called ``simple model of chemical evolution'' leads to implicit
integral equations that must be solved iteratively. We note in passing
that the dependence of the IMF on the metallicity is well established
theoretically, as low-metallicity gas cools less efficiently and
self-gravitating clumps are more resilient to fragmentation. This
leads to the formation of dense cores of higher mass. For this reason,
the IMF is supposed to be extremely biased towards massive stars if
the metallicity is smaller than $\sim$ 10$^{-4}$ Z$_\odot$ (Schneider
et al. 2002). Also the dependence of the IMF on the SFR is nowadays
observationally well established (Hoversten \& Glazebrook 2008; Meurer
et al. 2009; Gunawardhana et al. 2011; Kroupa 2014), and the IGIMF
theory is the only existing computable access accounting for these
observations (Weidner et al. 2013). Thus, in spite of the inherent
complexity, detailed simulations of galaxy evolution based on variable
IMFs need to be performed.  With this paper, we aimed at showing in a
simple setting how to take into account IMF variations in models of
the chemical evolution of galaxies (see also Martinelli \& Matteucci
2000; Calura et al. 2010).  The method outlined here can be used in
order to extend analytical studies of the evolution of galaxies based
on simple models of chemical evolution (e.g. Spitoni et al. 2010;
Lilly et al. 2013; Pipino et al. 2014).  We described how the solution
of the simple model differs from the standard, textbook analytical
solutions of the simple model. We also showed (building on previous
results of K\"oppen et al. 2007) that the IGIMF theory naturally leads
to a mass-metallicity relation. In fact, low-mass galaxies are
characterized on average by smaller SFRs.  According to the IGIMF
theory, a low SFR leads to a steep, top-light IMF, in which the
production of heavy elements by massive stars is extremely limited.
More massive galaxies instead produce many more massive stars because
of the higher level of SFR, hence the attained present-day metallicity
is larger.

We outline here once again that the main aim of the paper was not to
provide the ultimate explanation for the mass-metallicity relation.
Many detailed theoretical studies have been performed on this subject
(see e.g.  Matteucci \& Chiosi 1983; Tremonti et al.  2004; Finlator
\& Dav{\'e} 2008; Calura et al. 2009; Peeples \& Shankar 2011; Yin et
al. 2011 among many others).  As shown by Fig.  \ref{fig:mz}, the MZ
relation obtained with the IGIMF theory does not fit well the data and
assumptions about flow rates are required in order to obtain a good
fit.  Nevertheless it is important to show that $(i)$ The IGIMF theory
(as well as any other theory according to which the IMF is top-heavier
in more massive systems) naturally accounts for the MZ relation;
$(ii)$ Even within the IGIMF formalism, flow rates (infall and
outflows) remain crucial ingredients for the evolution of galaxies;
$(iii)$ The intensity of the galactic wind must be assumed to be
inversely proportional to the initial gaseous mass of the model
galaxy, in accordance with observations and theoretical studies.  The
outflow rates are however considerably smaller than the ones needed to
reproduce the MZ relation in models with invariant IMFs such as the
one of Spitoni et al. (2010).


In a companion paper this methodology will be applied to the study of
the mass-metallicity relation in dark-matter-free tidal dwarf
galaxies. At the same time, efforts are ongoing to model by means of
hydrodynamical codes the full chemo-dynamical evolution of galaxies
within the IGIMF theory, in order to assess the reliability of the
results obtained in this paper.

\section*{Acknowledgements} We thank an anonymous referee for
suggestions and constructive criticisms which considerably improved
the quality of the paper.  We thank E. Spitoni for providing us the
data reported in Fig. \ref{fig:lambda_comp}


\begin{thebibliography}{}

\bibitem[]{} Alves, J., Lombardi, M., \& Lada, C.~J.\ 2007, A\&A, 462,
  L17

\bibitem[]{} Banerjee, S., \& Kroupa, P.\ 2013, ApJ, 764, 29 

\bibitem[]{} Bastian, N., Covey, K.~R., \& Meyer, M.~R.\ 2010, ARA\&A,
  48, 339

\bibitem[]{} Bekki, K.\ 2013, MNRAS, 436, 2254

\bibitem[]{} Birrer, S., Lilly, S., Amara, A., Paranjape, A., \&
  Refregier, A.\ 2014, arXiv:1401.3162

\bibitem[]{} Calura, F., Pipino, A., Chiappini, C., Matteucci, F., \&
  Maiolino, R.\ 2009, A\&A, 504, 373

\bibitem[]{} Calura, F., Recchi, S., Matteucci, F., \& Kroupa, P.\
  2010, MNRAS, 406, 1985

\bibitem[]{} Cappellari, M., McDermid, R.~M., Alatalo, K., et al.\
  2012, Nature, 484, 485

\bibitem[]{} Conroy, C., \& van Dokkum, P.~G.\ 2012, ApJ, 760, 71 

\bibitem[]{} Dabringhausen, J., \& Kroupa, P.\ 2013, MNRAS, 429, 1858 

\bibitem[]{} Dabringhausen, J., Kroupa, P., \& Baumgardt, H.\ 2009,
  MNRAS, 394, 1529

\bibitem[]{} Dabringhausen, J., Kroupa, P., Pflamm-Altenburg, J., \&
  Mieske, S.\ 2012, ApJ, 747, 72

\bibitem[]{} Dayal, P., Ferrara, A., \& Dunlop, J.~S.\ 2013, MNRAS,
  430, 2891

\bibitem[]{} Dutton, A.~A., Treu, T., Brewer, B.~J., et al.\ 2013,
  MNRAS, 428, 3183

\bibitem[]{} Finlator, K., \& Dav{\'e}, R.\ 2008, MNRAS, 385, 2181 

\bibitem[]{} Gunawardhana, M.~L.~P., Hopkins, A.~M., Sharp, R.~G., et
  al.\ 2011, MNRAS, 415, 1647

\bibitem[]{} Hopkins, P.~F.\ 2013, MNRAS, 433, 170

\bibitem[]{} Hopkins, P.~F., Quataert, E., \& Murray, N.\ 2012, MNRAS,
  421, 3522

\bibitem[]{} Hoversten, E.~A., \& Glazebrook, K.\ 2008, ApJ, 675, 163

\bibitem[]{} Hsu, W.-H., Hartmann, L., Allen, L., et al.\ 2012, ApJ,
  752, 59

\bibitem[]{} Kirby, E.~N., Cohen, J.~G., Guhathakurta, P., et al.\
  2013, ApJ, 779, 102

\bibitem[]{} Kirk, H., \& Myers, P.~C.\ 2011, ApJ, 727, 64 

\bibitem[]{} K{\"o}ppen, J., Weidner, C., \& Kroupa, P.\ 2007, MNRAS,
  375, 673

\bibitem[]{} Kroupa, P.\ 2001, MNRAS, 322, 231

\bibitem[]{} Kroupa, P.\ 2002, Science, 295, 82

\bibitem[]{} Kroupa, P.\ 2014, arXiv:1406.4860

\bibitem[]{} Kroupa, P., \& Weidner, C. 2003, ApJ, 598, 1076

\bibitem[]{} Kroupa, P., Weidner, C., Pflamm-Altenburg, J., et al.\
  2013, Planets, Stars and Stellar Systems.~Volume 5: Galactic
  Structure and Stellar Populations, 115 (astro-ph/1112.3340)

\bibitem[]{} Lada, C.J., \& Lada, E.A. 2003, ARA\&A, 41, 57

\bibitem[]{} Lada, C.~J., Margulis, M., \& Dearborn, D.\ 1984, ApJ,
  285, 141

\bibitem[]{} Lara-L{\'o}pez, M.~A., Cepa, J., Bongiovanni, A., et al.\
  2010, A\&A, 521, L53

\bibitem[]{} Larson, R.~B.\ 1998, MNRAS, 301, 569

\bibitem[]{} Lee, H., Skillman, E.~D., Cannon, J.~M., et al.\ 2006,
  ApJ, 647, 970

\bibitem[]{} Lilly, S.~J., Carollo, C.~M., Pipino, A., Renzini, A., \&
  Peng, Y.\ 2013, ApJ, 772, 119

\bibitem[]{} Mac Low, M.-M., \& Ferrara, A.\ 1999, ApJ, 513, 142 

\bibitem[]{} Maeder, A.\ 1992, A\&A, 264, 105 

\bibitem[]{} Mannucci, F., Cresci, G., Maiolino, R., Marconi, A., \&
  Gnerucci, A.\ 2010, MNRAS, 408, 2115

\bibitem[]{} Marks, M., \& Kroupa, P.\ 2012, A\&A, 543, A8 

\bibitem[]{} Marks, M., Kroupa, P., Dabringhausen, J., \& Pawlowski,
  M.~S.\ 2012, MNRAS, 422, 2246

\bibitem[]{} Martin, C.~L.\ 2005, ApJ, 621, 227

\bibitem[]{} Martin, C.~L., Kobulnicky, H.~A., \& Heckman, T.~M.\
  2002, ApJ, 574, 663

\bibitem[]{} Martinelli, A., \& Matteucci, F.\ 2000, A\&A, 353, 269 

\bibitem[]{} Matteucci, F.\ 1994, A\&A, 288, 57 

\bibitem[]{} Matteucci, F.\ 2001, The Chemical Evolution of the
  Galaxy. Astrophysics and Space Science Library (Kluwer Academic
  Publishers)

\bibitem[]{} Matteucci, F., \& Chiosi, C.\ 1983, A\&A, 123, 121 

\bibitem[]{} Meurer, G.~R., Wong, O.~I., Kim, J.~H., et al.\ 2009,
  ApJ, 695, 765

\bibitem[]{} Ott, J., Walter, F., \& Brinks, E.\ 2005, MNRAS, 358,
  1453

\bibitem[]{} Padoan, P., \& Nordlund, {\AA}.\ 2002, ApJ, 576, 870 

\bibitem[]{} Pagel, B.~E.~J.\ 1997, Nucleosynthesis and Chemical
  Evolution of Galaxies, Cambridge University Press

\bibitem[]{} Peeples, M.~S., \& Shankar, F.\ 2011, MNRAS, 417, 2962 

\bibitem[]{} Pflamm-Altenburg, J., Gonz{\'a}lez-L{\'o}pezlira, R.~A.,
  \& Kroupa, P.\ 2013, MNRAS, 435, 2604

\bibitem[]{} Pflamm-Altenburg, J., \& Kroupa, P.\ 2009, ApJ, 706, 516

\bibitem[]{} Pflamm-Altenburg, J., Weidner, C., \& Kroupa, P.\ 2007,
  ApJ, 671, 1550

\bibitem[]{} Pipino, A., Lilly, S.~J., \& Carollo, C.~M.\ 2014,
  arXiv:1403.6146

\bibitem[]{} Pipino, A., \& Matteucci, F.\ 2004, MNRAS, 347, 968 

\bibitem[]{} Ploeckinger, S., Hensler, G., Recchi, S., Mitchell, N.,
  \& Kroupa, P.\ 2014, MNRAS, 437, 3980

\bibitem[]{} Puchwein, E., \& Springel, V.\ 2013, MNRAS, 428, 2966 

\bibitem[]{} Recchi, S., Calura, F., \& Kroupa, P.\ 2009, A\&A, 499,
  711

\bibitem[]{} Recchi, S., Calura, F., Gibson, B.~K., \& Kroupa, P.\
  2014, MNRAS, 437, 994

\bibitem[]{} Recchi, S., Matteucci, F., \& D'Ercole, A.\ 2001, MNRAS,
  322, 800

\bibitem[]{} Recchi, S., \& Hensler, G.\ 2013, A\&A, 551, A41 

\bibitem[]{} Recchi, S., Spitoni, E., Matteucci, F., \& Lanfranchi,
  G.~A.\ 2008, A\&A, 489, 555

\bibitem[]{} Schneider, R., Ferrara, A., Natarajan, P., \& Omukai, K.\
  2002, ApJ, 571, 30

\bibitem[]{} Spitoni, E., Calura, F., Matteucci, F., \& Recchi, S.\
  2010, A\&A, 514, A73

\bibitem[]{} Tinsley, B.~M.\ 1980, Fund. Cos. Phys., 5, 287

\bibitem[]{} Tremonti, C.~A., Heckman, T.~M., Kauffmann, G., et al.\
  2004, ApJ, 613, 898

\bibitem[]{} van den Hoek, L.~B., \& Groenewegen, M.~A.~T.\ 1997,
  A\&AS, 123, 305

\bibitem[]{} Weidner, C., \& Kroupa, P.\ 2004, MNRAS, 348, 187 

\bibitem[]{} Weidner, C., \& Kroupa, P. 2005, ApJ, 625, 754

\bibitem[]{} Weidner, C., \& Kroupa, P.\ 2006, MNRAS, 365, 1333 

\bibitem[]{} Weidner, C., Kroupa, P., \& Bonnell, I.~A.~D.\ 2010,
  MNRAS, 401, 275

\bibitem[]{} Weidner, C., Kroupa, P., \& Larsen, S.~S.\ 2004, MNRAS,
  350, 1503

\bibitem[]{} Weidner, C., Kroupa, P., Pflamm-Altenburg, J., \&
  Vazdekis, A.\ 2013, MNRAS, 436, 3309

\bibitem[]{} Woosley, S.~E., \& Weaver, T.~A.\ 1995, ApJS, 101, 181

\bibitem[]{} Yin, J., Matteucci, F., \& Vladilo, G.\ 2011, A\&A, 531,
  A136


\end{thebibliography}
\end{document}